# Energy Efficiency Testing and Modeling of a Commercial O-RAN System

White Paper

February 2026

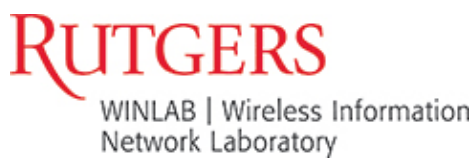 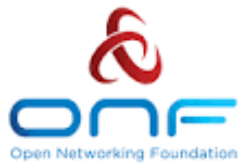 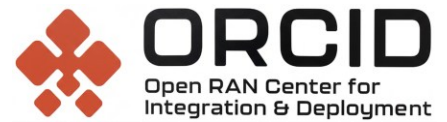

# Table of Contents

# Executive Summary

Network energy efficiency is of critical importance to mobile network operators for economic and ecological reasons. Therefore, reducing network energy consumption is a high priority. Beyond the use of more energy-efficient hardware components, operators increasingly rely on energy-saving features that dynamically adapt network configuration to reduce power consumption during periods of reduced traffic demand. The advent of the O-RAN architecture has brought disaggregation and virtualization, and in order to achieve the highest energy savings gains, we need rigorous measurement, analysis, and modeling of energy consumption at both the component and system levels. However, there remains a lack of publicly-available, quantitative data characterizing the behavior of commercial-grade O-RAN systems.

In this white paper, we present a detailed energy-efficiency characterization and modeling of a commercial O-RAN system based on comprehensive power and performance measurements, using a network deployment that faithfully replicates a production O-RAN network deployed by a wireless carrier.

The results are drawn from an energy test campaign conducted through a joint collaboration between the Open RAN Center for Integration and Deployment (ORCID) Lab Testing and Evaluation (T&E) Project and the ONF / Rutgers WINLAB Energy Efficiency Research and Development (R&D) project, both supported by the first round (NOFO-1) of the U.S. National Telecommunications and Information Administration (NTIA) Public Wireless Supply Chain Innovation Fund (PWSCIF) grants . The test environment includes an O-RAN system with an Amazon Web Services-hosted O-CU, a dedicated-server O-DU, and six high-power, multi-band O-RUs representative of commercial macro-cell network deployments.

Our results identify the dominant factors influencing power consumption across the O-RAN stack and quantify energy usage variation under different operational and traffic scenarios. Based on these findings, we provide practical recommendations for lightweight additions to existing equipment test procedures that enable routine energy measurement. These measurements can be used by operators to parameterize power-consumption models, ultimately supporting data-driven energy optimization and more sustainable operation of commercial O-RAN networks.

# Acknowledgements

This white paper is a result of an excellent collaboration between the ONF/WINLAB Energy Efficiency R&D Project and the ORCID Lab T&E Project. We appreciate the support and funding of the NTIA Public Wireless Supply Chain Innovation Fund. The team members are:

**Rutgers WINLAB**: N.K. Shankaranarayanan, Zhuohuan Li, Ivan Seskar, Prasanthi Maddala
**ONF/Aether:** Sarat Puthenpura, Alex Stancu
**ORCID**: Jens Sohn, Akash Gupta, Sreenidhi Parthasarathy, Wilfred Luiz, Jeffrey Williamson, VenkataReddy Varra

For further information, please contact N. K. Shankaranarayanan (shankar@winlab.rutgers.edu).

# 1. Introduction

With the densification of base stations and the use of advanced antenna technologies like Multiple-Input Multiple-Output (MIMO), the energy demand of 5G networks has become a critical concern. An important part of the 5G network evolution is the concept of Open Radio Access Network (O-RAN), an open, disaggregated, and software defined architecture that enables multi-vendor interoperability, flexible network deployments, and integration of artificial intelligence (AI) for dynamic network management. However, despite these benefits, the energy consumption of O-RAN based 5G networks is still a crucial issue due to the increased complexity introduced by the disaggregation and virtualization of network functions, which are both key features of O-RAN which have a major impact on energy efficiency. The energy profile characteristics of physical network functions (such as O-RU), virtualized and cloudified network functions (such as O-CU, O-DU, Core) and the supporting infrastructure (O-Cloud) are all different. Understanding and profiling the power consumption characteristics and the interplay under various network and traffic conditions is thus particularly important.

There are several ongoing discussions to standardize tests and features in the Energy Efficiency/Energy Savings area in the O-RAN Alliance. Energy efficiency has emerged as a critical Key Performance Indicator (KPI) for next-generation RAN architectures [1], with approximately 73% of the network consumption attributed to widely distributed base stations [2]. This scenario poses a substantial threat to the sustainability of wireless networks, underscoring the urgent need for innovative energy-saving solutions. The global deployment of networks based on the 5G O-RAN architecture presents an energy efficiency paradox: while the disaggregated architecture with O-RUs enhances network flexibility and promotes multi-vendor interoperability, the lack of standardized energy efficiency test methods and metrics for components and end-to-end systems adds to the complexity of power management.

Energy efficiency in O-RAN is influenced by a range of factors, including hardware technology efficiency, optimization of radio access network (RAN) resources, and the intelligent management of network traffic. A key opportunity in O-RAN energy optimization lies in reducing the power consumption of O-RUs while maintaining high-quality service levels. Since energy optimization has multiple dimensions, it is critical to have a good methodology for O-RAN energy testing, and also a good power consumption prediction model to be used in network optimization when resources can be scaled back and/or devices turned off to save energy. While there have been studies and models of power consumption of O-RUs and pre-O-RAN base stations, there are very few detailed measurements available in the literature.

In this white paper, we describe the energy efficiency of a commercial O-RAN system, with detailed measurements of power consumption, and models of energy efficiency. We present results of an energy test campaign at the Open RAN Center for Integration and Deployment (ORCID) lab performed as part of a joint collaboration of the ORCID lab project and the ONF/WINLAB Energy Efficiency R&D project, which are both projects funded by the first round (NOFO-1) of the U.S. National Telecommunications and Information Administration (NTIA) Public Wireless Supply Chain Innovation Fund (PWSCIF) grants. The results include power and performance measurements in the ORCID O-RAN system with an AWS-cloud based O-CU and a dedicated-server O-DU supporting six high-power, multi-band O-RUs. The results identify the key factors influencing the power consumption and provide insights into how the power consumption changes for different test scenarios. We also provide recommendations for simple additions of energy measurements to equipment tests which can be used by operators to parametrize the power consumption models which can drive energy optimization.

Our collaboration was between our university-based Test R&D project (Energy Efficiency) and a commercial T&E Lab (ORCID), and thus a perfect complement of capabilities. It is an example of maximum leverage of NTIA research funding to meet the NTIA PWSCIF objectives.

The objective of the collaboration between ONF, WINLAB, and ORCID Lab was to:

- Share and validate energy testing methodology from the POET testbed [3]
- Gather more data from a commercial system to augment research testbed results

- Validate and expand power consumption models
- Develop insights into test methods, and operator perspectives on energy savings

It is important to emphasize that it was not a vendor comparison or evaluation exercise. The system was tested on an as-is basis. The reported results are based on the system as deployed and represent a snapshot in time of the technological trend. We expect the performance of systems to change. The objective was to validate methodology and gain insights to develop and quantify energy efficiency models. As a research exercise, it was also not an effort to develop detailed energy conformance test specifications at this stage, but the results would be of relevance to specifications in [4].

Section 2 provides a high-level system description. The Test Cases and the Test Methodology are presented in sections 3 and 4. Section 5 has the results and model for the O-RU power consumption. Section 6 and 7 have the results of the O-DU and O-CU power consumption. Section 8 summarizes the energy efficiency.

# 2. System Description

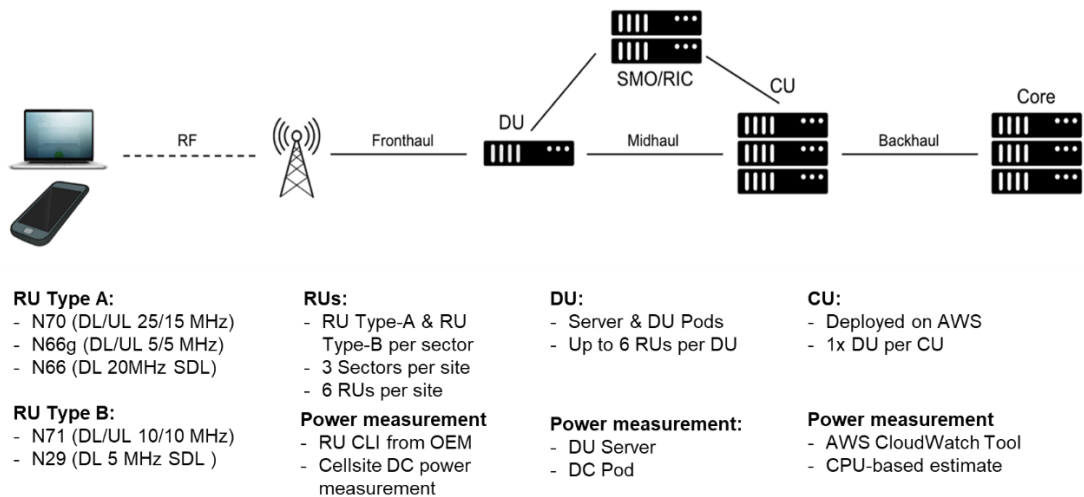

Figure 1. Energy test setup in the end-to-end test line at the ORCID Lab

The ORCID Lab provided a commercial O-RAN system test line aligned with actual deployments of a nationwide, commercial-grade O-RAN network. In contrast to other research labs and open labs, this provides an end-to-end system designed and provisioned with operational considerations driving the choice of components and configurations. Of particular importance is the use of two types (type A and B) of multi-band O-RUs, and an O-DU instance which can support six O-RUs (three each of type A and type B). As part of the project, the test systems were augmented to include power consumption measurements. Figure 1 shows the energy test setup in the ORCID Lab. More details are in later figures and sections.

Figure 2 shows a block diagram of the end-to-end O-RAN system with a 5G Core, 1 O-CU, 1 O-DU, and 6 O-RUs for a three sector site, with one type-A RU and one type-B RU for each sector. Totally, there are up to 12 cells (with 2 cells/carriers per RU). There are 30 dB attenuators at the output of the RU antenna ports and the RF path loss through the patch panel is approximately 20 dB. The signal quality is excellent thus allowing full throughput at the highest MCS level. There is a common Cell Site Router (CSR) connecting to all the O-RUs. There is also SiteBoss equipment that can measure the DC power consumed by each of the O-RUs. In addition, there is a Command Line Interface (CLI) to all the RUs.

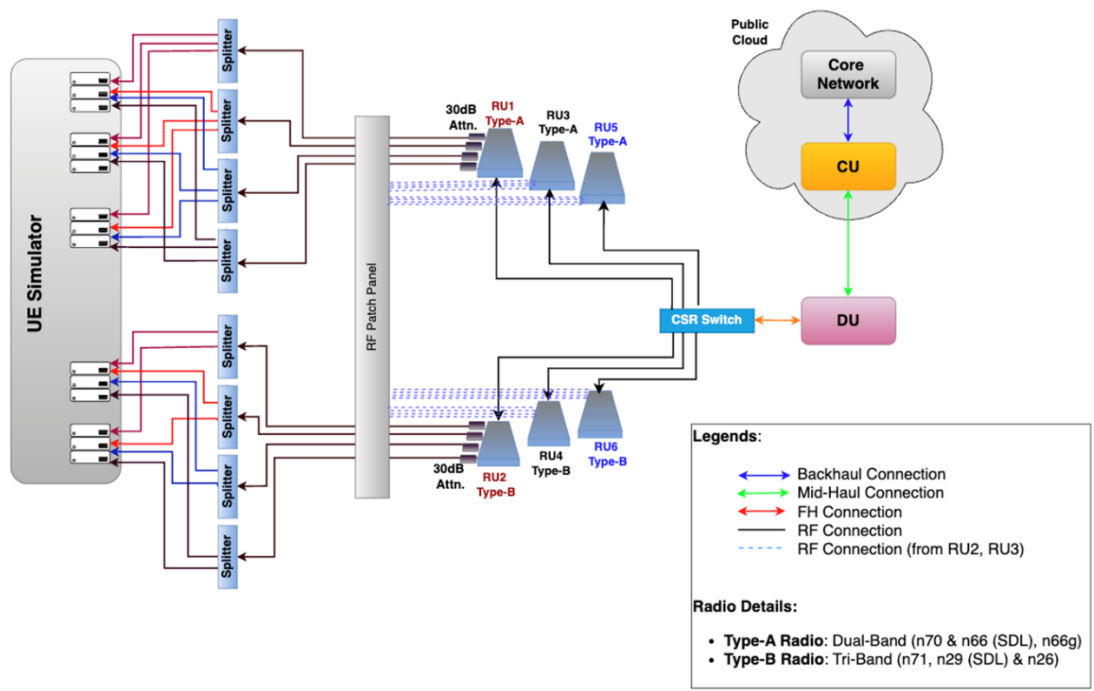

Figure 2. Block diagram of an end-to-end O-RAN system with a three sector site, with one type-A RU and one type-B RU for each sector.

Figure 3 shows the details of the multi-band FDD Type-A and Type-B O-RUs, with the details of the band configurations shown in Figure 4. The Type-A O-RU supports two FDD bands, each with its RF stage and a set of 4 power amplifiers. The second n66 band supports two different carriers. Thus, there are a total of 3 carriers and each of them could go up to 46 dBm per antenna. This means that the n66 stage and 4 power amplifiers are supporting twice the power. The Type-B O-RU supports three FDD bands, but only the first two (n71, n29) were used in our tests and the n26 band was not used. The Type-A RU n66 carrier and the Type-B RU n29 carrier are both Secondary Downlink (SDL) carriers and need to be used with a primary carrier. The Type-B RU n29 SDL carrier and the Type-A RU n66 SDL carrier were used with the Type-B O-RU n71 carrier as the primary carrier (in different test cases).

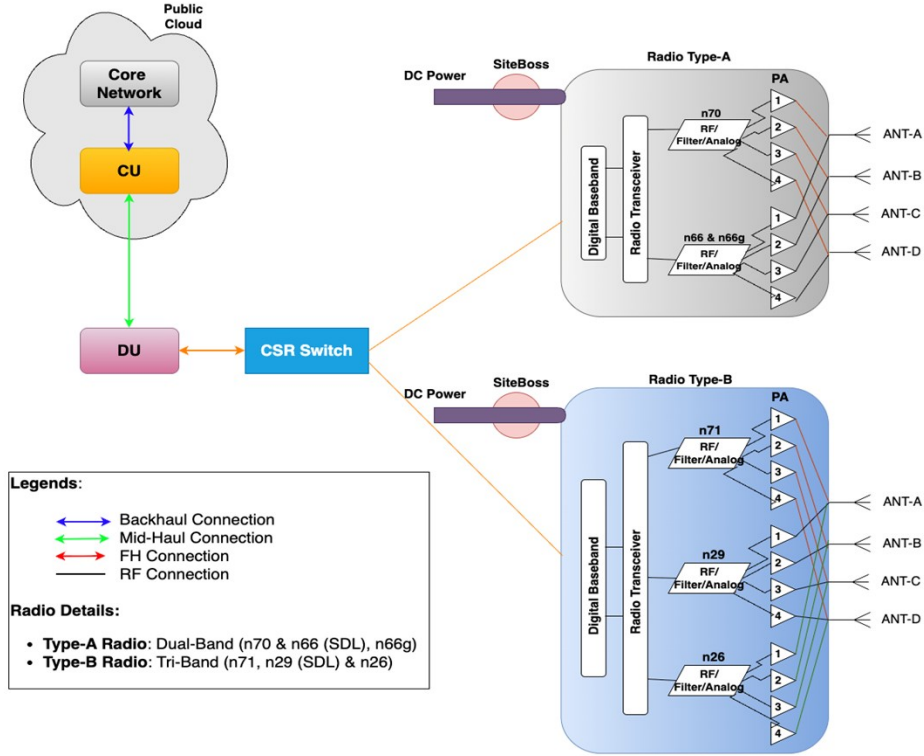

Figure 3. Block diagram of O-RAN test system showing details of the Type-A and Type-B O-RUs. Each O-RU supports multiple FDD bands.

| Radio Type | Band | Layers | Channel Bandwidth | | Throughput (Mbps) | |
|---|---|---|---|---|---|---|
| | | | DL | UL | DL | UL |
| Type A | n70 | 4 | 25 | 15 | 440 | 57 |
| | n66 | 4 | 20 | 0 | 350 | 0 |
| | n66g | 4 | 5 | 5 | 62 | 16 |
| Type B | n71 | 2 | 10 | 10 | 70 | 37 |
| | n29 | 2 | 5 | 0 | 40 | 0 |

Note:
1) Type B RU n26 band not used in this set of test cases
2) n66 and n29 can only be used with Carrier Aggregation (CA)

Figure 4. Band configuration for the Type A and Type B O-RUs. The n66 and n29 carriers are Secondary Downlink (SDL) carriers and need to be used with a primary carrier. Both the 29 SDL carrier and the n66 SDL carrier were used with the n71 carrier as the primary carrier, in different test cases (TC 25,27,29,30 in section 3).

# 3. Test Cases

We constructed several test cases to understand how the energy consumption of the O-system varied with different scenarios. The objective is to (a) determine the key factors, (b) identify the small number of key tests to be included in an energy test specification, and (c) develop a model which can be used to predict the power consumption. The possible variations are:

- RF Transmit power level (Tx gain)
- Number of MIMO layers
- Full and partial downlink throughput
- TCP v UDP traffic
- Number of carriers (per RU) carrying data
- Number of RUs carrying data

| TC. No. | Radio Type | Traffic Type | Traffic Load (%) DL | Traffic Load (%) UL | Traffic Direction | # of UEs | Channel Bandwidth (MHz) DL | Channel Bandwidth (MHz) UL | MIMO | Gain Value (dBm) | # of Cells | Bands |
|---|---|---|---|---|---|---|---|---|---|---|---|---|
| 1/2 | Type-A | TCP/UDP | 100% | 100% | DL & UL | 64 | 25 | 15 | 4 x 4 | 37 | 1 | n70 |
| 3 | Type-A | TCP | 100% | 100% | DL & UL | 64 | 25 | 15 | 4 x 4 | 40 | 1 | n70 |
| 4 | Type-A | TCP | 100% | 100% | DL & UL | 64 | 25 | 15 | 4 x 4 | 43 | 1 | n70 |
| 5 | Type-A | TCP | 100% | 100% | DL & UL | 64 | 25 | 15 | 4 x 4 | 46 | 1 | n70 |
| 6 | Type-A | TCP | 100% | 100% | DL | 64 | 25 | 15 | 4 x 4 | 37 | 1 | n70 |
| 7 | Type-A | TCP | 100% | 100% | DL & UL | 64 | 25 | 15 | 2 x 2 | 37 | 1 | n70 |
| 8 | Type-A | TCP | 100% | 100% | DL & UL | 64 | 25 | 15 | 2 x 2 | 43 | 1 | n70 |
| 17/18 | Type-A | TCP/UDP | 30% | 100% | DL & UL | 64 | 25 | 15 | 4 x 4 | 37 | 1 | n70 |
| 19/20 | Type-A | TCP/UDP | 50% | 100% | DL & UL | 64 | 25 | 15 | 4 x 4 | 37 | 1 | n70 |
| 9/10 | Type-A | TCP/UDP | 100% | 100% | DL & UL | 64 | 5 | 5 | 4 x 4 | 37 | 1 | n66g |
| 11/12 | Type-B | TCP/UDP | 100% | 100% | DL & UL | 64 | 10 | 10 | 2 x 2 | 37 | 1 | n71 |
| 13 | Type-B | TCP | 100% | 100% | DL | 64 | 10 | 10 | 2 x 2 | 37 | 1 | n71 |

Figure 5. Single-carrier test cases where only one carrier has downlink data. Each colored cell highlights the difference in the test case compared to previous test cases. For example, Test Case 3 (TC 3) has 40 dBm RF gain value where TC 1 and 2 have 37 dBm. Test cases with TCP and UDP variations of traffic are shown together (e.g., TC1/2, TC 17/18).

Figure 5 lists the details of single-carrier test cases, and Figure 6 lists details of multi-carrier/multi-RU test cases. Additional test cases were executed in December 2025 after initial test cases to add more information. TC11 & TC12 were executed again to verify the power consumption of RU2 in 2x2 MIMO mode with other two RF chains fully off to confirm the extra power consumption for active, unused RF chains. TC 25 & TC27 were executed again to confirm that DU power consumption does not show large variations during data download due to external factors. TC 31 & TC 32 included higher powers for RU1 n66g. TC 33 & TC 34 were variants of TC27 (low pathloss, MCS 27) using medium path loss (MCS 14) and high path loss (MCS 4) respectively.

| TC. No. | Radio Type | Traffic Type | Traffic Load (%) | | Traffic Direction | # of UEs | Channel Bandwidth (MHz) | | | | | | | | MIMO | | Gain Value (dBm) | # of Cells | # of Sectors |
|---|---|---|---|---|---|---|---|---|---|---|---|---|---|---|---|---|---|---|---|
| | | | | | | | Radio Type-A | | | | | Radio Type-B | | | | | | | |
| | | | | | | | n70 | | n66 | n66g | | n71 | | n29 | Radio Type-A | Radio Type-B | | | |
| | | | DL | UL | | | DL | UL | DL | DL | UL | DL | UL | DL | | | | | |
| 14 | Type-A | TCP | 100% | 100% | DL & UL | 64 | 25 | 15 | | 5 | 5 | | | | 4 x 4 | 2 x 2 | 37 | 2 | 1 |
| 15/16 | Type-A & Type-B | TCP/UDP | 100% | 100% | DL & UL | 64 | 25 | 15 | | 5 | 5 | 10 | 10 | | 4 x 4 | 2 x 2 | 37 | 3 | 1 |
| 25 | Type-A & Type-B | TCP | 50% | 100% | DL & UL | 32 | 25 | 15 | | 5 | 5 | 10* | 10 | 5* | 4 x 4 | 2 x 2 | 37 | 4 | 3 |
| 27 | Type-A & Type-B | TCP | 100% | 100% | DL & UL | 32 | 25 | 15 | | 5 | 5 | 10* | 10 | 5* | 4 x 4 | 2 x 2 | 37 | 4 | 3 |
| 29/30 | Type-A & Type-B | TCP/UDP | 100% | 100% | DL & UL | 32 | | | 20* | 5 | 5 | 10* | 10 | | 4 x 4 | 2 x 2 | 37 | 3 | 1 |

Figure 6. Multi-carrier test cases where there is data on more than one carrier. In most of the test cases, both Type-A and Type-B RU are carrying traffic.

The salient features of each test case are listed below:

- TC 1/2: Baseline test with TCP (TC1) and UDP (TC2) for a single n70 carrier with 4x4 MIMO and 100% downlink traffic
- TC3, TC4, TC5: Increasing values of RF Transmit gain compared to TC1
- TC6: Downlink only version of TC1
- TC7 and TC 8: 2x2 MIMO version of TC1 and TC4
- TC 17/18 and TC 19/20: 30% and 50% downlink traffic compared to TC 1/2 based on [5]
- TC 9/10: Similar to TC1/2 but with n66g carrier
- TC 11/12: Similar to TC1/2 but with Type-B RU n71 carrier
- TC 13: Downlink-only version of TCP 11, similar to TC 6
- TC 14: Two-carrier version of TC1 with both n70 and n66g carriers
- TC 15/16: TCP/UDP tests for two O-RUS, one 2-carrier Type-A (n70, n66g) and one 1-carrier Type B (n71)
- TC 27: TCP test for six O-RUs, three 2-carrier Type-A (n70, n66g) and three 2-carrier Type B (n71,n29) with n29 being secondary carrier for n71
- TC 25: Variant of TC27 with 50% downlink data
- TC 29/30: TCP/UDP tests for six O-RUs, three 2-carrier Type-A (n66, n66g) and three 1-carrier Type B (n71) with n66 being secondary carrier for n71
- TC 31, TC32: variant of TC 9/10 with higher transmit gains (43 dBm, 46 dBm).
- TC 33: variant of TC1 with medium pathloss resulting in 225 Mb/s DL throughput (51% of TC1 440 Mb/s) with MCS 15, RSRP −95 dBm, using 64QAM modulation
- TC 34: variant of TC1 with high pathloss resulting in 75 Mb/s DL throughput (51% of TC1 440 Mb/s with MCS 4, RSRP −110 dBm, using 95% QPSK and 5% 16QAM modulation

# 4. Power & Performance Measurement Techniques

The O-RUs had a command line interface (CLI) which provided metrics reported by the O-RU. During this test, the O1 interface was not in place. However, the metrics available using the CLI are representative of the metrics that would be available over O1. The OEM CLI metrics included:
- RF Tx gain setting (per carrier, per antenna)
- RF Tx power estimate - per antenna, per carrier. The RF Tx power was typically within 0.5 to 1 dB of the Tx gain setting.
- Power consumption estimate

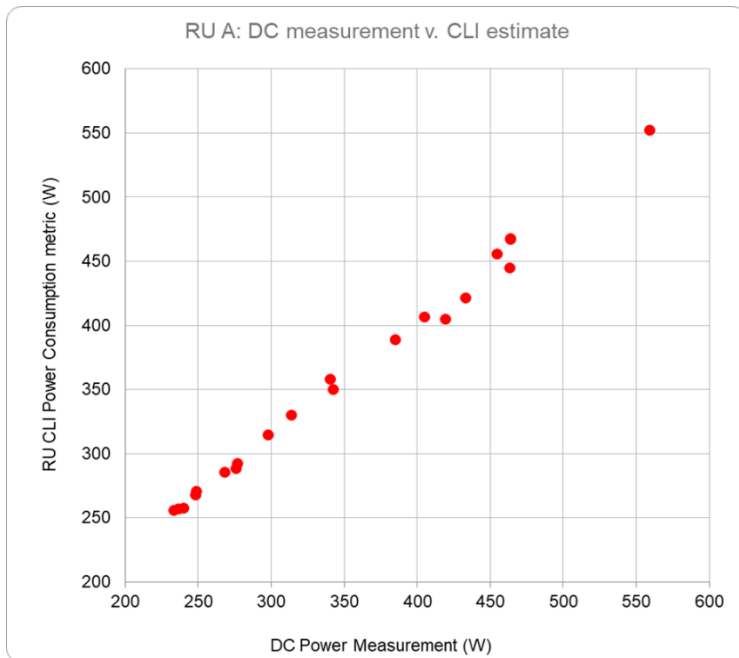

Figure 7. Scatter plot showing correlation of self-reported O-RU power consumption (Y-axis) with external power measurement (X-axis).

The O-RU power consumption was measured using three methods:
- Clamp meter (measuring current to O-RU). This was instrumented for one O-RU and was deemed unsuitable since it was hard to replicate for all O-RUs
- Cellsite (SiteBoss) DC power measurement. This is a ground-truth external measurement of power consumed. This had a granularity of 0.1 A current and about a 5@ granularity of power given the -54V voltage.
- O-RU CLI OEM Tool: The CLI tool reported the power consumption every second.

The cell site external measurement (SiteBoss) was selected to be the reference (although precision is only 5W). Since an O-RAN operator system would get the self-reported power metric over O1, we compared the two methods. Figure 7 shows a scatter plot of the CLI metric (y-axis) with the external DC power measurement (x-axis). They are correlated well as expected and are within +/- 10%. About 5% of the error could be due to the 5W granularity. For the results, we used the external DC power measurement for the O-RU power consumption.

The end-to-end throughput measurement was available in the UE Simulator.

# 5. Power Consumption in multi-band O-RUs

Given the opportunity to use and study multi-band O-RUs, we included test cases which would enable us to understand the power consumption and further refine our model.

## 5.1. Multi-band O-RU Power Consumption: Test Results

We used different test cases covering the activation of only one band/carrier in the O-RU at a time as well as the activation of two bands in the O-RU at the same time (test cases TC1, TC9, TC14, TC29.). Such tests reveal any difference in efficiencies of the RF stages/amplifiers for the two bands and also allow us to parametrize the O-RU power consumption model.

Figure 8 shows the O-RU power consumption v. total RF power for RU Type-A for a single N70 carrier, for different Tx gain levels, for both 4x4 MIMO and 2x2 MIMO configurations (test cases TC1, TC3, TC4, TC5, TC7, TC8). The Tx Gain value is shown in parenthesis () for each data point. It also shows similar measurements for a single N66g carrier (4x4 MIMO). The RF power is extracted from the CLI (Command Line interface) tool values and are within 0.5 to 1 dB of the Tx Gain setting. For both 2x2 and 4x4 MIMO, for 0 RF power, there is a substantial idle power consumption of about 200 W. We can see that the O-RU power consumption increases with RF power at high RF powers in a near-linear manner. For the 43 dBm and 37 dBm gain settings, the total RF power for 2x2 MIMO is half of the total RF power for 4x4 MIMO. This is expected since the 2x2 MIMO uses only two of the antennas and power amplifiers instead of 4. Note that the power consumption does not drop to half due to the high idle power consumption. Also, the power consumption for 4x4 MIMO / 40 dBm gain is similar, but slightly higher than 2 x 2 MIMO / 43 dBm, with the same total RF power. Figure 8 also shows the results for the N66g carrier which has a higher idle power as well as a steeper slope. A steeper slope indicates a lower power amplifier efficiency.

Figure 9 compares the results for the single-carrier (N70 or N66g) and dual-carrier (both N70 and N66g) operation for RU Type-A. We can clearly see the increase in the idle power consumption for dual-carrier activation, and that the combined amplifier efficiency is a blend of the efficiencies of the two RF stages.

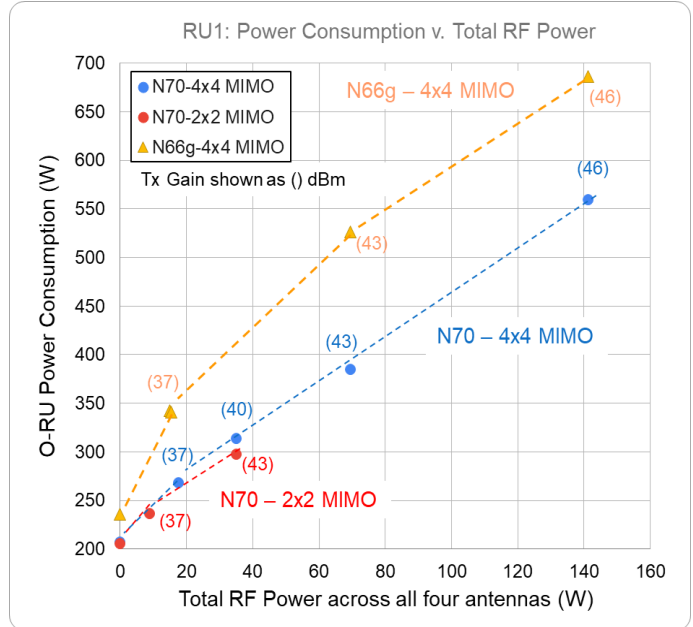

Figure 8. O-RU power consumption v total RF power for RU Type-A for a single N70 carrier (4x4 MIMO and 2x2 MIMO) and a single N66g carrier (4x4 MIMO), for different Tx gain levels. The Tx gain value is shown in parenthesis () for each data point. The slopes of the lines indicate the power amplifier efficiency (steeper is less efficient).

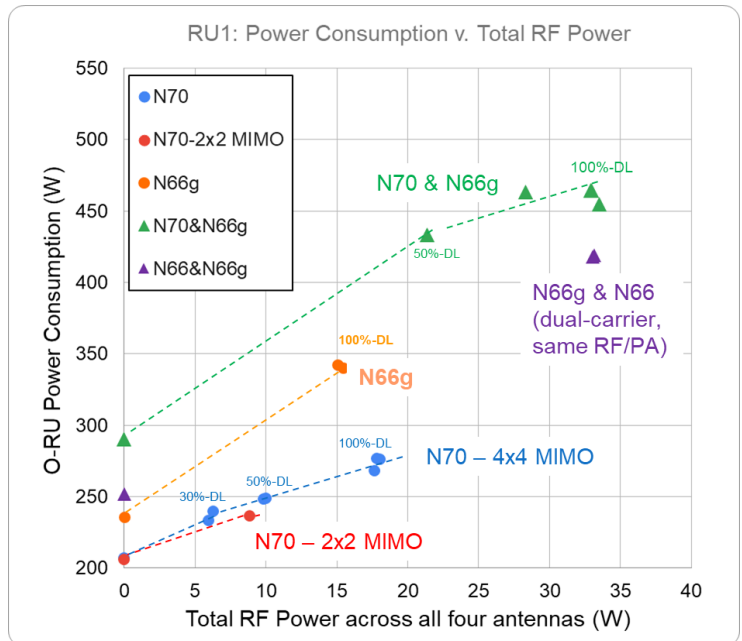

Figure 9. O-RU power consumption v total RF power for RU Type-A for a single (N70 only - blue, N66g only-orange) and multiple (N70 & N66g - green, N66 & N66g - purple) carriers. The Tx gain is 37 dBm.

## 5.2. Multi-band O-RU Power Consumption Model

Based on Figure 9, we can expand the O-RU power consumption model described in [6]. Since there are multiple bands, each with their own RF processing stage and set of power amplifiers (PA), we modify the model equation to sum over the multiple RF modules. This is shown in Figure 10.

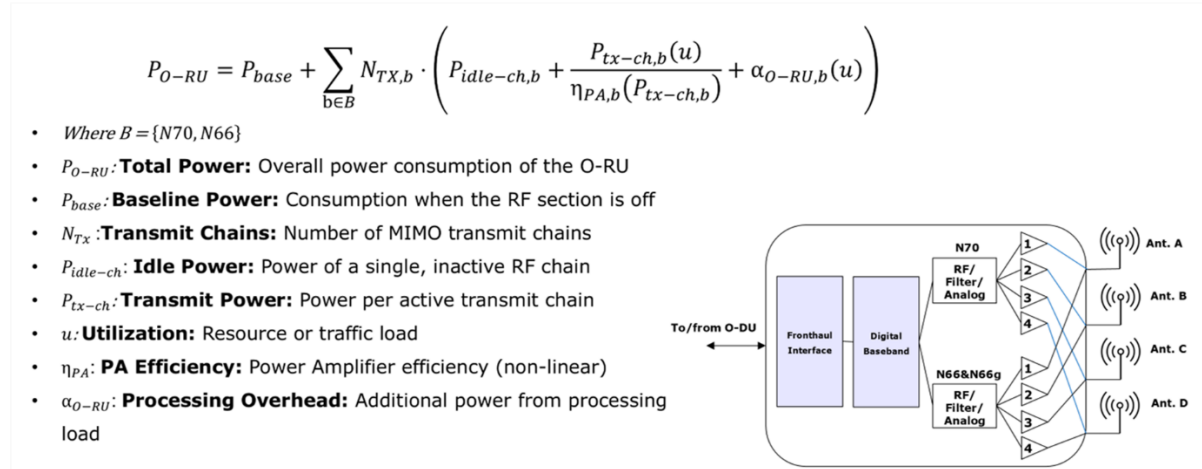

$$P_{O-RU} = P_{base} + \sum_{b \in B} N_{TX,b} \cdot \left( P_{idle-ch,b} + \frac{P_{tx-ch,b}(u)}{\eta_{PA,b}(P_{tx-ch,b})} + \alpha_{O-RU,b}(u) \right)$$

- Where $B = \{N70, N66\}$
- $P_{O-RU}$: **Total Power:** Overall power consumption of the O-RU
- $P_{base}$: **Baseline Power:** Consumption when the RF section is off
- $N_{Tx}$: **Transmit Chains:** Number of MIMO transmit chains
- $P_{idle-ch}$: **Idle Power:** Power of a single, inactive RF chain
- $P_{tx-ch}$: **Transmit Power:** Power per active transmit chain
- $u$: **Utilization:** Resource or traffic load
- $\eta_{PA}$: **PA Efficiency:** Power Amplifier efficiency (non-linear)
- $\alpha_{O-RU}$: **Processing Overhead:** Additional power from processing load

Figure 10. Multi-band O-RU Power Consumption Model.

The multiple test cases in section 5.1 allow us to estimate the parameters in the above model. For example, we can write:

$$P_{O-RU,idle} = P_{base} + \text{sum} ( P_{idle-ch,b} \text{ from active RF chains} )$$

From the results of Test Case 1, 9, 14, we have the power consumption when there is no data transmission (idle), and we can thus write:

$$207 \text{ W} = P_{base} + P_{idle-ch,N70}$$
$$236 \text{ W} = P_{base} + P_{idle-ch,N66g}$$
$$291 \text{ W} = P_{base} + P_{idle-ch,N70} + P_{idle-ch,N66g}$$
$$\Rightarrow P_{idle-ch,N70} = 291 \text{ W} - 236 \text{ W} = 55 \text{ W}$$
$$\Rightarrow P_{idle-ch,N66g} = 291 \text{ W} - 207 \text{ W} = 84 \text{ W}$$
$$\Rightarrow P_{base} = 152 \text{ W}$$

We can estimate the Power Amplifier efficiency of each of the bands when only that band is active:

$$\eta_{PA} \approx \frac{P_{out}}{P_{O-RU} - P_{idle}} = \frac{N_{TX} \cdot P_{tx-ch}}{P_{O-RU} - P_{idle}}$$

We can estimate that the N70 band power amplifier ranges from 29% (37 dBm) to 33% (40 dBm) to 39% (43 dBm). This PA had a 43 dBm max rating. The N66g band PA had a max rating of 48 dBm. For the N66g band PA, we estimate the power amplifier efficiency ranged from 14% (37 dBm) to 32% (46 dBm). These power consumption estimates are depicted in Figure 11. The N66g RF stage is designed to handle two carriers, and the power amplifiers have high power ratings as required for operation in the band. This results in higher power consumption for that stage.

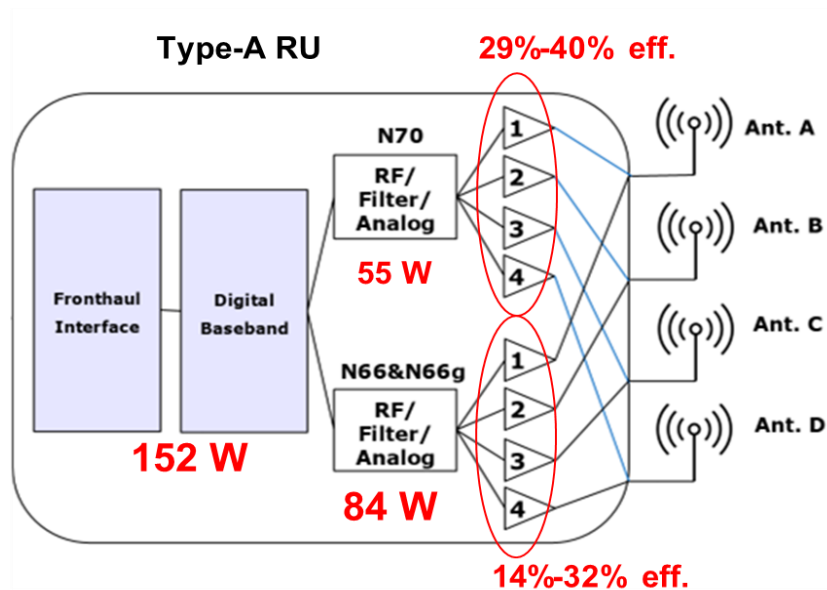

Figure 11. Type-A O-RU power model shown with parameters derived by simple tests.

We do not have the same detailed tests for the Type-B O-RU. There are tests for only N71 and joint N71 & N29, but we do not have only N29 (SDL). The Type-B O-RU has 4 antennas, but the ORCID Lab configuration was to use it in a 2x2 MIMO mode based on operational considerations. The test cases call for the Type-B O-RU to be in 2x2 mode with only 2 of the 4 RF streams/antennas enabled. While executing the test case TC11, the team did a test with the 3rd and 4th RF stage/antenna active but not carrying data and this increased the idle power contribution by about 30 W.

Thus, we can parametrize the RU power model with simple tests which can be added to energy test specification. The Power Model is especially useful to predict power consumption; for example, we saw that there was only 1% error for predicting power consumption for 2x2 MIMO and for the 50% throughput.

# 6. O-DU Power Consumption

The O-DU in the ORCID Lab setup is hosted on a Dell XR11 server with a Kubernetes infrastructure supplied by an O-Cloud vendor, and an O-DU Network Function from an O-DU vendor. The DU includes an Intel ACC100 accelerator card. The DU instance is provisioned to support six O-RUs (3 Type-A and 3 Type-B). The test cases included full-buffer max downlink throughput scenarios ranging from: (a) only one RU, one carrier to (b) only one RU, two-carriers to c) two RUs (1-sector configuration) to (d) six RUs (3-sector configuration). As the O-DU supported more RUs and more throughput, the power consumption showed some increase, due to the extra processing. The power increase in proportion to the idle baseline is relatively modest, and most noticeable for the maximum load condition.

The O-DU power consumption was measured in two ways. One was the total server power consumption and this is equivalent to methods such as BMC, Redfish, or Dell IDRAC. In addition, the O-Cloud system also reports an estimate of the total power consumed by the pods used for the O-DU NF and the PTP functionality. Figure 12 shows the power consumption of the server and the DU/PTP pods as a function of time when the data traffic is switched on for the 2-RU and 6-RU case. The total server power consumption is about 130W higher than the DU/PTP pods. This is remarkably similar to the measurements in the WINLAB POET servers and BubbleRAN Kubernetes deployment [3].

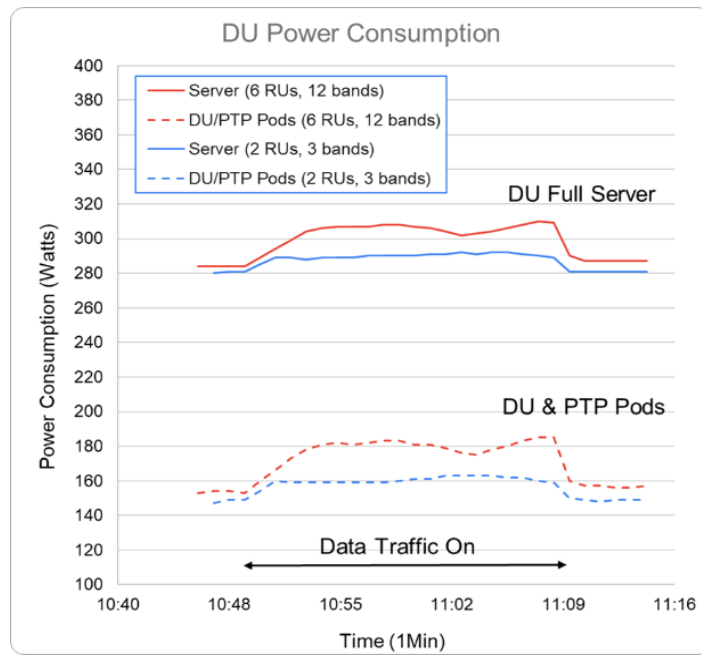

Figure 12. Power consumption of O-DU shown for the full DU server, and also for only the sub-portion of the power consumed by the Kubernetes pods used for the DU and the PTP support. The gap of about 130 W represents power consumption (e.g., server fan) not visible to the O-Cloud system (e.g., Kubernetes).

An O-Cloud deployment has visibility into the O-DU pods. It may not have visibility of the total server power consumption. Depending on the operator deployment model, all of the total server power consumption may be attributed in total to the O-RAN power consumption. For example, in the case of a dedicated server for the O-RAN system and the DU, the total server power consumption should be included in the power consumption. Figure 13 shows the detail of the total DU server power consumption. Compared to an idle state, there is an increase of about 25

W for 1836 Mb/s (using 6 RUs) and about 10 W for 572 Mb/s (using 2 RUs). The idle power (no data) for 6 RUs is about 5 W higher. Compared to the total O-RAN system power (see section 8), the power consumption increase in the DU during data transmission is modest.

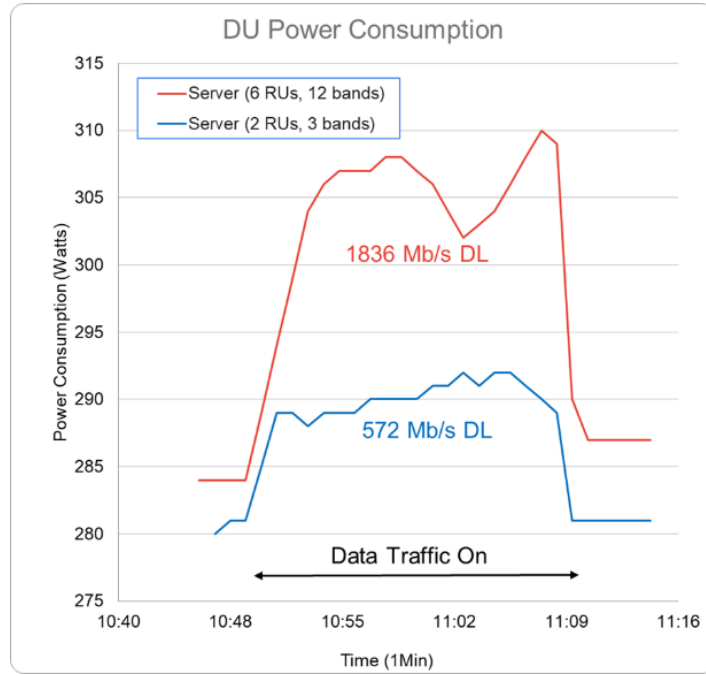

Figure 13. Details of power consumption of total O-DU server showing about 10% increase during downlink data transmission of 1836 Mb/s and about 5% increase for 572 Mb/s.

# 7. O-CU Power Consumption

In the ORCID Lab setup, the O-CU and O-DU network functions are provided by the same vendor. Unlike the cell site Dell server, the O-CU is deployed on third-party AWS Cloud. The O-CU is provisioned to support 17 DUs and in this case, it is supporting only one DU. The team studied AWS Cloudwatch tools as well as other Kubernetes tools for estimating the power consumption of the CUCP and CUUP pods. The conclusion was to use the AWS Cloudwatch tool which reports the CPU utilization of the various containers making up the CUUP and CUCP NFs. Figure 14 shows screenshots of the AWS Cloudwatch tool for the CUUP (showing the pod responsible for data traffic) and the CUCP (showing ue-connection-manager pod).

The CPU utilization during idle and during data transmission can be monitored. There is an estimate from AWS for the maximum power consumption rating for the server being used. There is also an estimate for the server power consumption when the server CPU utilization is 0. The team used the following method to map the CPU utilization to a power estimate of the CU pods.

$$P = P_{idle} + ( P_{max} - P_{idle} ) \times U/100$$
P = Estimated CPU power consumption (Watts)
Pidle = Power when CPU utilization = 0% (Watts)
Pmax = Power when CPU utilization = 100% (Watts)
U = Current CPU utilization (%)

There was a very small increase in CPU utilization when the data download was active. The resultant increase in power consumption was only about 1 W. This low number is perhaps due to the fact that the CU was provisioned to handle 17 times the data and the workload here is relatively small.

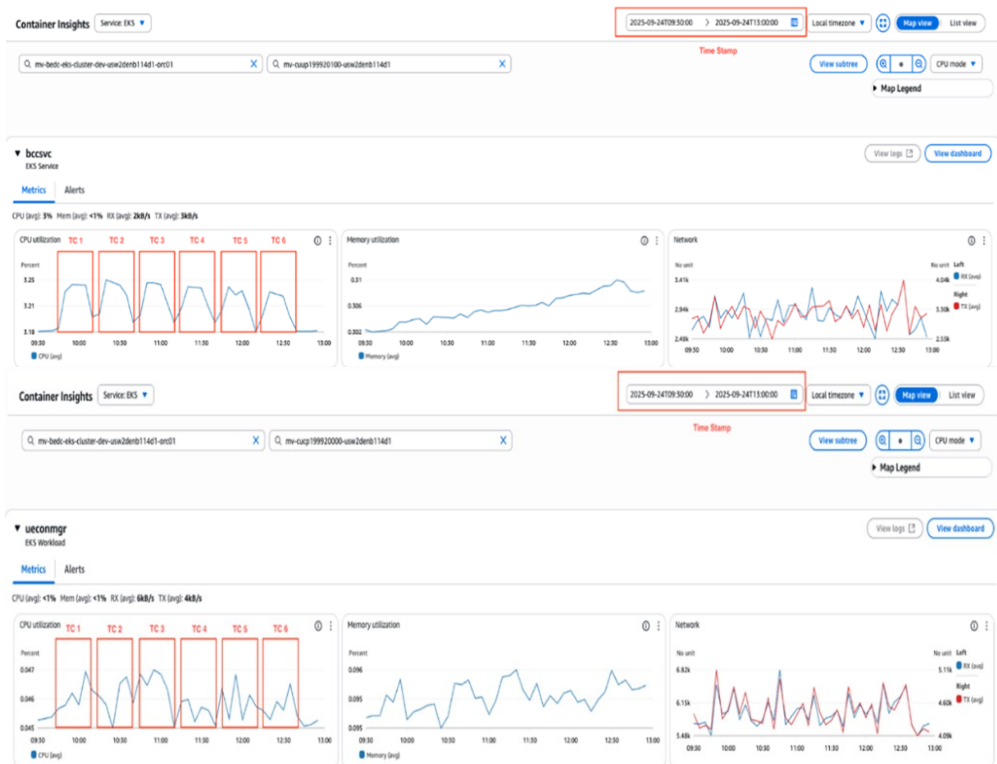

Figure 14. AWS Cloudwatch tool screenshots showing detail of power consumption of CUUP and CUUP pods of the O-CU server

# 8. Energy Efficiency Results

The test cases were designed to measure the power consumption with and without data traffic. The throughput was provided by the UE Simulators, while the power consumption was provided by the methods described earlier. We are interested in energy efficiency, which is a measure of how well electrical power being consumed is used to deliver a resource or value to the end user. In general,

$$\text{Network energy efficiency} = \{\text{Measure of desired outcome/value}\} / \{\text{Electrical energy consumed}\}$$

For an end-to-end 5G system, one of the prevalent measures for energy efficiency is based on PDCP SDU data volume as per 3GPP TS 28.554, clause 6.7.1 which is given by:

$$EE_{MN,DV} = (\text{Total SDU data volume (bits)}) / (\text{Total energy consumed by participating network elements (Joules)})$$

For the downlink, this is equivalent to:

$$EE_{MN,DV} = (\text{Total downlink throughput (bits/sec)}) / (\text{Total power consumed by participating network elements (Watts)})$$

We note that the above system energy efficiency does not capture the effect of lower throughput due to reduced MCS values for high path loss between the RU and UE. The ETSI Dynamic Energy Test [7] uses 3 UE groups with 3 path loss values to incorporate such effects. In most of our test cases, we had low path loss and all UEs had the highest MCS. We also repeated some test cases for medium and high path loss.

For the O-RU itself, the objective of the O-RU is to faithfully deliver RF radio power with the information provided by the DU, and we can write:

$$\text{O-RU efficiency} = \text{Total RF Power} / \text{RU\_Power}$$
$$\text{O-RU PA efficiency} = \text{RF Power} / (\text{RU\_Power} - \text{Idle\_Power})$$

Figure 15 shows results for the system energy efficiency for selected test cases. For each test case, we have the downlink throughput (per-band, total per-RU, total system), RF transmit power per RU, power consumption (per-RU, total RU, DU, CU, total system), and data throughput energy efficiency in kbps/W (ratio of total downlink throughput to total system power consumption). The low loss conditions ensure max throughput (MCS=27) and throughput is constrained by the bandwidth. For RU type-A with 4x4 MIMO, we have 440 Mb/s and 62 Mb/s for the 25 MHz (N70) and 5 (N66g) MHz bandwidth carriers respectively. For RU type-B with 2x2 MIMO, we have 70 Mb/s and 40 Mb/s for the 10 MHz (N71) and 5 MHz (N29) bandwidth carriers.

The highest energy efficiency of 655 kb/s/W is for full 100% downlink of 1836 Mb/s for test case TC27, the 3-sector configuration of 6 RUs (with full config of two active bands per RU). This is the highest throughput. The highest channel bandwidth is the N70 25 MHz carrier. The second-highest energy efficiency of 563 kb/s/W is for the single N70-only carrier test case TC1. **The takeaway is that: **given the overhead of the idle power consumption (without data), we see the best energy efficiency is when we can have high throughputs using the high bandwidth channels.**

| No | Test Case Notes | RU1 - DL TP (Mb/s) 70 | 66g | 66* | Total | RU2 - DL TP (Mb/s) 71 | 29* | Total | Total DL (Mb/s) | RF Power (W) RU1 | RU2 | Total | Power Consumption (W) RU1 | RU2 | All RUs | DU | CU | Total | Energy Eff (kbps/W) |
|---|---|---|---|---|---|---|---|---|---|---|---|---|---|---|---|---|---|---|---|
| 1 | RU-1, 25MHz | 440 | | | 440 | | | | 440 | 18 | | 18 | 268 | | 268 | 284 | 230 | 782 | 563 |
| 19 | RU-1, 25MHz - 50% DL | 220 | | | 220 | | | | 220 | 10 | | 10 | 248 | | 248 | 285 | 230 | 763 | 288 |
| 17 | RU-1, 25MHz - 30% DL | 132 | | | 132 | | | | 132 | 6 | | 6 | 234 | | 234 | 282 | 230 | 746 | 177 |
| 7 | RU-1, 25MHz - 2 x 2 MIMO | 220 | | | 220 | | | | 220 | 9 | | 9 | 236 | | 236 | 282 | 230 | 748 | 294 |
| 9 | RU-1, 5MHz | | 62 | | 62 | | | | 62 | 15 | | 15 | 343 | | 343 | 283 | 230 | 855 | 73 |
| 14 | RU-1, (25MHz+5MHz) | 440 | 62 | | 502 | | | | 502 | 33 | | 33 | 464 | | 464 | 287 | 230 | 981 | 512 |
| 11 | RU-2, 10 MHz | | | | | 70 | | 70 | 70 | | 8 | 8 | | 212 | 212 | 283 | 230 | 725 | 97 |
| 15 | RU-1 (25MHz+5MHz) & RU-2 (10) | 440 | 62 | | 502 | 70 | | 70 | 572 | 33 | 8 | 41 | 455 | 218 | 673 | 289 | 230 | 1192 | 480 |
| 29 | RU-1 (20MHz+5MHz) & RU-2 (10) | | 62 | 350 | 412 | 70 | | 70 | 482 | 33 | 7 | 40 | 418 | 200 | 618 | 283 | 230 | 1131 | 426 |
| 27 | RU-1 (25MHz+5MHz) & RU-2 (10MHz+5MHz) - 3S | 440 | 62 | | 502 | 70 | 40 | 110 | 1836 | 28 | 16 | 44 | 463 | 292 | 2267 | 304 | 230 | 2802 | 655 |
| 25 | RU-1 (25MHz+5MHz) & RU-2 (10MHz+5MHz) -3S, 50% DL | 220 | 31 | | 251 | 35 | 20 | 55 | 918 | 21 | 13.5 | 35 | 433 | 279 | 2138 | 299 | 230 | 2666 | 344 |

Figure 15 Energy Efficiency Results from selected Test Cases. For each test case, we have the downlink throughput (per-band, total per-RU, total system), RF power per RU, power consumption (per-RU, total RU, DU, CU, total system), and data throughput energy efficiency.

We now focus on certain test cases to understand the trends. Figure 16 focuses on test cases comparing single carrier and multi-carrier configuration.

| No | Test Case Notes | RU1 - DL TP (Mb/s) 70 | 66g | 66* | Total | RU2 - DL TP (Mb/s) 71 | 29* | Total | Total DL (Mb/s) | RF Power (W) RU1 | RU2 | Total | Power Consumption (W) RU1 | RU2 | All RUs | DU | CU | Total | Energy Eff (kbps/W) |
|---|---|---|---|---|---|---|---|---|---|---|---|---|---|---|---|---|---|---|---|
| 1 | RU-1, 25MHz | 440 | | | 440 | | | | 440 | 18 | | 18 | 268 | | 268 | 284 | 230 | 782 | 563 |
| 9 | RU-1, 5MHz | | 62 | | 62 | | | | 62 | 15 | | 15 | 343 | | 343 | 283 | 230 | 855 | 73 |
| 14 | RU-1, (25MHz+5MHz) | 440 | 62 | | 502 | | | | 502 | 33 | | 33 | 464 | | 464 | 287 | 230 | 981 | 512 |

Figure 16 Energy efficiency for single-carrier and multi-carrier configuration.

TC1 with the 25 MHz N70 carrier is the most efficient single-carrier configuration. It achieves 440 Mb/s and 563 kbps/W efficiency. The N66g band with 5 MHz bandwidth achieves only 62 Mb/s and 73 kbps/W efficiency, and it is not efficient to use the N66g band by itself. However, activating the 5 MHz carrier in addition to the existing 25 MHz carrier achieves 502 Mb/s and 512 kbps/W efficiency. Given the RU power consumption overhead already being "spent" with the N70 carrier operation, it is worth adding a second carrier to the RU as opposed to activating a second RU. The bit rate increases by 14% and total power increases by 25%.

Next, we compare the single-sector and multi-sector configuration where the DU supports more RUs and a higher throughput. Figure 17 focuses on TC14,15,27. Growing from 1 sector (2 cells/sector) to 3 sectors (4 cells/sector), which also means growing from total 1 RU (2 cells) to total 6 RUs (total 12 cells) improves energy efficiency. The total DL bit rate increases from 502

Mb/s to 1836 Mb/s which is a 3.65x increase. The total power consumption increases from 981 W to 2802 W which is a 2.86x increase. The energy efficiency thus improves from 512 to 655 kbps/W which is a 28% increase. **The takeaway is that energy efficiency improves when DU and CU overhead is amortized over more RUs and cells.**

| No | Test Case Notes | RU1 - DL TP (Mb/s) 70 | 66g | 66* | Total | RU2 - DL TP (Mb/s) 71 | 29* | Total | Total DL (Mb/s) | RF Power (W) RU1 | RU2 | Total | Power Consumption (W) RU1 | RU2 | All RUs | DU | CU | Total | Energy Eff (kbps/W) |
|---|---|---|---|---|---|---|---|---|---|---|---|---|---|---|---|---|---|---|---|
| 14 | RU-1, (25MHz+5MHz) | 440 | 62 | | 502 | | | | 502 | 33 | | 33 | 464 | | 464 | 287 | 230 | 981 | 512 |
| 15 | RU-1 (25MHz+5MHz) & RU-2 (10) | 440 | 62 | | 502 | 70 | | 70 | 572 | 33 | 8 | 41 | 455 | 218 | 673 | 289 | 230 | 1192 | 480 |
| 27 | RU-1 (25MHz+5MHz) & RU-2 (10MHz+5MHz) - 3S | 440 | 62 | | 502 | 70 | 40 | 110 | 1836 | 28 | 16 | 44 | 463 | 292 | 2267 | 304 | 230 | 2802 | 655 |

Figure 17 Energy efficiency for single sector and multi-sector configuration

In an operational system, the throughput will not always be at 100% maximum levels. The throughput can be lower for two reasons: (a) medium and high path loss can reduce the channel quality, SIR, and MCS values, thus causing lower throughput. (b) the total demand from all users may be less than the maximum capacity of the system. We executed test cases to study the second scenario where UE demand is for partial capacity. This is the spirit of the ETSI static energy test [5] where the downlink traffic is adjusted to be at 50% and 30% of the maximum. Figure 18 focuses on the downlink throughput variations. This was achieved by constraining the available PRB resources in the channels to a 50% or 30% level. This automatically caused the UE Simulator to achieve proportionately less throughput.

| No | Test Case Notes | RU1 - DL TP (Mb/s) 70 | 66g | 66* | Total | RU2 - DL TP (Mb/s) 71 | 29* | Total | Total DL (Mb/s) | RF Power (W) RU1 | RU2 | Total | Power Consumption (W) RU1 | RU2 | All RUs | DU | CU | Total | Energy Eff (kbps/W) |
|---|---|---|---|---|---|---|---|---|---|---|---|---|---|---|---|---|---|---|---|
| 1 | RU-1, 25MHz | 440 | | | 440 | | | | 440 | 18 | | 18 | 268 | | 268 | 284 | 230 | 782 | 563 |
| 19 | RU-1, 25MHz - 50% DL | 220 | | | 220 | | | | 220 | 10 | | 10 | 248 | | 248 | 285 | 230 | 763 | 288 |
| 17 | RU-1, 25MHz - 30% DL | 132 | | | 132 | | | | 132 | 6 | | 6 | 234 | | 234 | 282 | 230 | 746 | 177 |
| 27 | RU-1 (25MHz+5MHz) & RU-2 (10MHz+5MHz) - 3S | 440 | 62 | | 502 | 70 | 40 | 110 | 1836 | 28 | 16 | 44 | 463 | 292 | 2267 | 304 | 230 | 2802 | 655 |
| 25 | RU-1 (25MHz+5MHz) & RU-2 (10MHz+5MHz) -3S, 50% DL | 220 | 31 | | 251 | 35 | 20 | 55 | 918 | 21 | 13.5 | 35 | 433 | 279 | 2138 | 299 | 230 | 2666 | 344 |

Figure 18 Energy efficiency for different downlink throughput values

The results for the single N70 carrier show that reducing traffic from 100% to 50% to 30% (using PRB assignment) reduces from 440 Mb/s to 220 Mb/s to 132 Mb/s as expected. The RU Tx RF Power drops from 18 W to 10 W to 6 W, but the RU power consumption drop from 268 W to 248 W to 23 W is relatively modest. Total system power consumption drops from 782W to 763 to 746 W which are only 2.5% and 4.7% drops for the 50% and 30% throughputs respectively. The energy efficiency drops from 563 kbps/W to 288 kbps/W to 177 kbps/W - which are 49% and 68% reduction. We see similar behavior for the 3-sector configuration with 6 RUs. Reducing traffic from 100% to 50% reduces DL throughput from 1836 Mb/s to 918 Mb/s. The total power consumption drops from 2802 W to 2666 W – only a 4.9% drop. The energy efficiency drops from 655 kbps/W to 344 kbps/W which is a 48% reduction. **The takeaway is that: with high overheads of RU and DU idle power, and without energy savings features enabled, a reduction in DL throughput reduces energy efficiency dramatically. It is best to operate carriers at high utilization levels for the best energy efficiency.**

| Test Case | | RU1 - DL TP (Mb/s) | | | | RU2 - DL TP (Mb/s) | | | Total DL (Mb/s) | RF Power (W) | | | Power Consumption (W) | | | | | | Energy Eff (kbps/W) |
|---|---|---|---|---|---|---|---|---|---|---|---|---|---|---|---|---|---|---|---|
| No | Notes | 70 | 66g | 66* | Total | 71 | 29* | Total | | RU1 | RU2 | Total | RU1 | RU2 | All RUs | DU | CU | Total | |
| 1 | RU-1, 25MHz | 440 | | | 440 | | | | 440 | 18 | | 18 | 268 | | 268 | 284 | 230 | 782 | 563 |
| 7 | RU-1, 25MHz - 2 x 2 MIMO | 220 | | | 220 | | | | 220 | 9 | | 9 | 236 | | 236 | 282 | 230 | 748 | 294 |

Figure 19. Energy efficiency comparison for 4x4 and 2x2 MIMO for single-carrier.

We also compared the energy performance when 4x4 MIMO is reduced to 2x2 MIMO mode. The results are shown in Figure 19 for a single N70 carrier. When we change from 4x4 MIMO to 2x2 MIMO, the DL bitrate reduces from 440 Mb/s to 220 Mb/s – 50% drop expected. The RU RF Power reduces from 18W to 9W since only 2 out of 4 antennas are used. The RU Power Consumption reduces from 268W to 236W – which is a 12% reduction (about 200 W is the overhead of idle power). The DU Power Consumption drop is only 2W. The total power consumption drops from 782W to 748 W which is a 4.4% drop. The energy efficiency drops from 563 kbps/W to 294 kbps/W which is a 48% reduction. **The takeaway is that: with the high overheads of RU idle power and DU overhead, it is best to achieve higher bitrate and 4x4 MIMO is better for energy efficiency.** Note that we do see a power consumption drop in the MIMO RU from 268 W to 236 W. If this were a higher-power 64 MIMO system changing to 32 MIMO, this change would be more significant and there might be operational reasons to adapt the massive MIMO to save energy as the throughput demand changes.

Finally, we also wanted to measure the throughput and power consumption for the default (low) path loss as well as medium and high path loss. As expected, the MCS and throughput values are lower, but there is no change in the power consumption. The results for RU-1, N70 carrier, 37 dBm Tx Gain, 4x4 MIMO are shown in Figure 20.

| TC. No. | Path loss | UE RSRP | DL MCS | Modulation | DL Bit rate | Total power (W) | Energy Efficiency (kbps/W) |
|---|---|---|---|---|---|---|---|
| 1 | Low | -65 dBm & -75 dBm | 27 | 64-QAM | 440 Mb/s | 782 W | 563 |
| 33 | Medium | -95 dBm | 15 | 64-QAM | 225 Mb/s | 782 W | 288 |
| 34 | High | -110 dBm | 4 | 95% QPSK and 5% 16QAM | 75 Mb/s | 782 W | 96 |

Figure 20. Energy efficiency comparison for n70 carrier (37 dBm Tx Gain, 4x4 MIMO) for low/medium/high pathloss conditions with different MCS and TCP downlink throughput values.

# 9. Conclusion

To our knowledge, we are presenting the first comprehensive quantitative results for power consumption and energy efficiency results for an end-to-end commercial O-RAN system with an O-DU supporting six high-power multi-band O-RUs. We have provided detailed results for a wide variety of test cases. Our results show the impact of dominant factors on power consumption: the baseline overhead of a component with zero traffic, the total RF power transmitted from the O-RU, and the power amplifier efficiency. We have validated and extended our previous O-RU model to predict power consumption for a multi-band O-RU. We show the influence of having an O-DU support multiple O-RUs. Finally, we have quantitative results for energy efficiency for a range of scenarios.

Energy efficiency is critical for the sustainable deployment of 5G and future wireless networks. The disaggregated nature of O-RAN presents both challenges and opportunities. Through this project, we have demonstrated that rigorous, standardized testing in a commercial setting, combined with validated power models can provide the foundation needed for effective energy optimization.

The knowledge, tools, and methodologies developed through this work contribute to the broader goals of sustainable wireless network deployment and provide actionable insights for the O-RAN community and mobile network operators worldwide. The comprehensive data sets, validated models, and test methodologies from this project will continue to serve the research community and industry as O-RAN technology evolves.

# 10. References

[1] O-RAN.WG1.NESUC-R003-v01.00.04: "O-RAN Working Group 1; Network Energy Saving Use Cases Technical Report

[2] NGMN Alliance, "Green Future Networks: Network Energy Efficiency v1.1, December 2021, www.ngmn.org

[3] N Shankaranarayanan, Zhuohuan Li, Ivan Seskar, Prasanthi Maddala, Sarat Puthenpura, Alexandru Stancu, "POET: A Platform for O-RAN Energy Efficiency Testing," 2024 IEEE Vehicular Technology Conference (VTC2024-Fall), Washington DC, USA, October 2024

[4] O-RAN.TIFG.E2E-Test.0-R003-v06.00: "O-RAN Test and Integration Focus Group, End-to-end Test Specification", https://oranalliance.atlassian.net/wiki/download/attachments/3096477976/O-RAN.TIFG.E2E-Test.0-R003-v06.00.docx?api=v2

[5] ETSI ES 202 706-1 Environmental Engineering (EE); Measurement method for energy efficiency of wireless access network equipment; Part 1: Power consumption - static measurement method (Note: Annex E covers NR)

[6] Zhuohuan Li, Prasanthi Maddala, N Shankaranarayanan, Ivan Seskar, Sarat Puthenpura, Alexandru Stancu, Christian Nunez Alvarez, Gregg Albrecht, "Energy Efficiency Testing and Power Modeling of O-RAN Radio Units," 2025 IEEE Future Networks World Forum (FNWF), Bangalore, India, 2025, pp. 1-6.

[7] ETSI TS 103 786 Environmental Engineering (EE); Measurement method for energy efficiency of wireless access network equipment Dynamic energy performance measurement method of 5G Base Station (BS). (Note: Annex C defined traffic mix)